\documentstyle[epsfig,12pt]{article}
\begin{document}
\vskip 1cm
\centerline{\bf STUDY OF THE AVALANCHE TO STREAMER TRANSITION}
\centerline{\bf IN GLASS RPC EXCITED BY UV LIGHT.}
\vskip 1cm
\centerline{Ammosov~V., Gapienko~V.,Kulemzin~A., 
Semak~A.,Sviridov~Yu.,Zaets~V.}
\vskip 0.5cm
\centerline{Institute for High Energy Physics, Protvino, Russia}

\begin{abstract}

A small glass RPC filled  with Ar/Isob./Freon mixture has been exposed
to a UV laser light. Avalanche  and streamer regimes of discharge were
reached in a fixed region of the RPC exited by the UV. A dependence of
avalanche-to-streamer transition process  on the laser  beam intensity
and on the high voltage  applied to the RPC was  studied. Two types of
the  streamer signal   have  been observed.  Using a  CCD  TV  camera,
pictures on multi-streamer propagation over RPC were obtained.

\end{abstract}

\section*{\bf Introduction}
\par Working voltages  for the RPC operated in  the avalanche mode  is
close  to values where avalanche  transforms into the streamer. If one
wants to use RPC in the proportional regime  he should be ready to see
from time to time the strong streamer signal. At  the low threshold of
electronics  any big  pulse  is not   desirable  because it produce  a
multichannel firing.  
\par 
An  attempt to get better understanding  on
how streamer grows from the avalanche was done  in the present work. A
glass RPC with a $2~mm$ gas gap was exposed to  the UV laser beam. The
UV laser allow to  fix  a start point of    growth and provide   equal
conditions  for  any  new avalanche.  By  variation  of  the  UV light
intensity we  tried to understand how  appearance of  the streamer and
its  characteristics depends on a  charge in the  avalanche and on the
working voltage  ($HV$).  Both: induced  signal from pick-up electrode
and light emitted by discharge in the RPC were registered.  

\par
Early a good correspondence between electrically induced signals and a light
emission from   the  glass   RPC was   found  in    \cite{OSAKA1}. The
correspondence was observed in following aspects: signal shapes, pulse
heights and  timing.  In \cite{OSAKA2}   one can  see pictures of  the
streamer  discharge in the glass RPC  taken using  image intensifier
and the CCD camera. In \cite{sem} was investigated a avalanche to streamer 
transition and presented pictures of a correspondent process. An estimation of 
a size of the space propagation of streamer and 
avalanche discharges was done. It  was mentioned about 
possible mechanism of the discharge in the RPC. In the present work a 
multi-streamer process were investigated. And laser help as to understand a
complex picture of this phenomena.

\section{\bf Experimental set-up}
\begin{figure}
\epsfxsize14cm\epsfysize12cm\epsffile{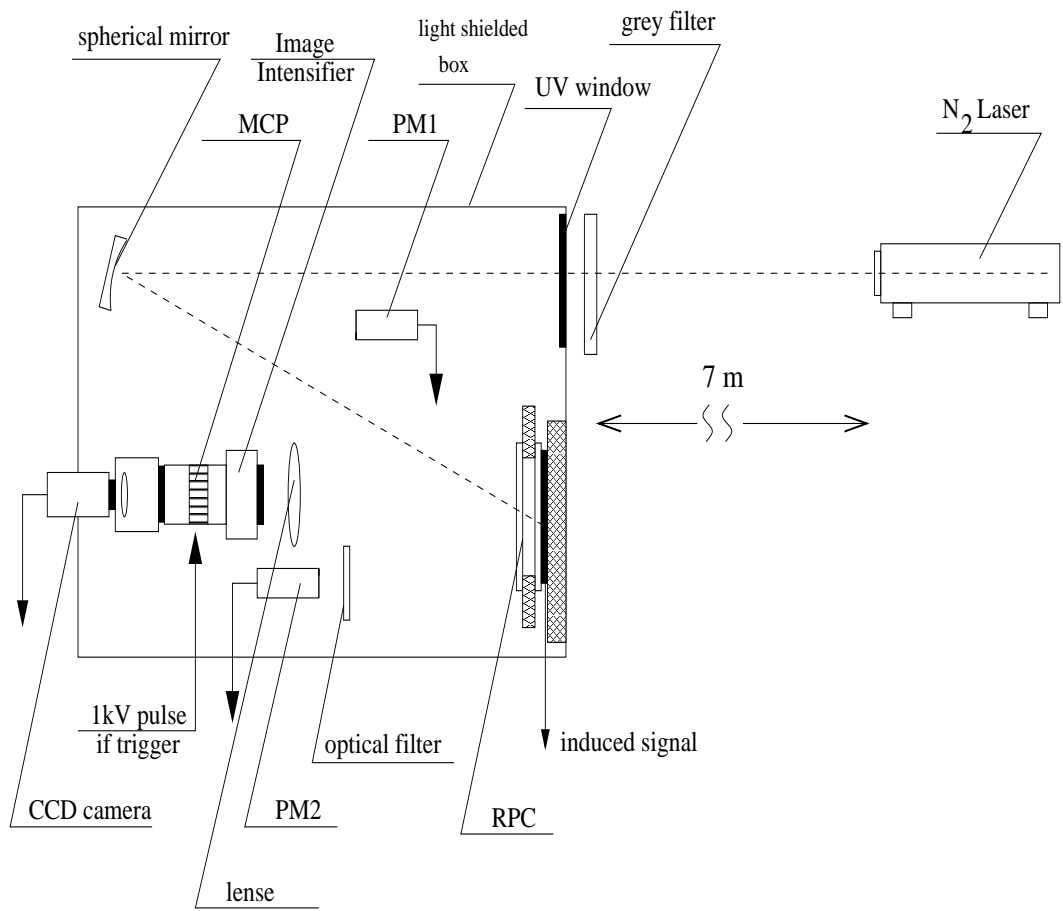}
\caption{\label{1} Experimental Set-up.}
\end{figure}

Fig.\ref{1}   shows experimental set-up.  A LGI-21   pulse laser was a
source of the  UV with $337~nm$ wave  length. A duration  of the laser
pulse  was   $8~ns$ with  a  maximal  energy  is about  $40~\mu  J$. A
frequency of light pulses can   be changed. Taking into account  $\sim
1~s$ recovery time  of our RPC  when  it was operated in  the streamer
mode  we chose the  pulse frequency so, that the   time between of two
discharges  in the RPC was  bigger than  $10~s$.  Being a  source of a
high-level  noise laser was installed  outside of  the test room $7~m$
far  from the RPC. The  UV beam was focused in  a $0.3~mm$ spot on the
entry glass electrode of  the  RPC by  a long-focus spherical   mirror
(SM).

\par
At first stage of the present work two photomultipliers (PM1, PM2) were used for monitoring of the laser beam intensity and for to measure spectrum of the light from discharges in the RPC. One of photomultipliers, PM1, was faced to the spherical mirror. This one with UV filter was used to measure part of the light scattered off SM because of mirror imperfection and thus for monitoring of a laser intensity. 
The second photomultiplier, PM2, faced to the RPC was used for a light registration from discharges occurring in the gas. By changing of optical filters between PM2 and RPC it was possible to measure different ranges of a discharge light spectrum.  Signals from both photomultipliers were fed to QDCs.
\par
RPC and PMs were housed in a light shielded box as it is shown in fig.\ref{1}. The laser UV entered this box through a window made of glass filter transparent for UV only.
The laser beam intensity ($I$) was changed by installation of grey optical filters between the laser and  the shielded box entry window.
\par 
At the final stage of the work, second photomultiplier (PM2) was replaced with a image intensifier and TV CCD camera to get picture of a streamer discharge distribution over the RPC area. 
\par
Our $10\times 10~cm^2$ RPC with a $2~mm$ gas gap was made of $2~mm$ glass plates. A bulk resistivity of glass was found to be $\sim 5\times 10^{12}\Omega \cdot cm$. A high voltage cathode was performed by  metalization of the glass plate surface. A transparency for the visible light of the  thin metallic layer was about 70\%. A 2~mm in diameter area in the center of the cathode was free of metalization. The UV beam entered the RPC through this area. 
\par 
An anode electrode was done with a layer of a carbon paint on the glass surface. And $2\times 2~mm$ free of paint area was in the center of the anode to the UV light can leave the RPC without producing a scintillation.       
\par
An induced signal from the anode was amplified by a U33 amplifier having a gain=$20$ and $400~MHz$ bandwidth. RPC signal and signals from photomultipliers were digitized with a 11-bits P267 module ("SUMMA" standard). The input sensitivity of the P267 is $\sim 0.3~pC/count$. When it was necessary to estimate the RPC efficiency, a discriminator with the $6~mV$ threshold was used after the amplifier.
\par
In the our study we felt RPC with $Ar/iso-C_{4}H_{10}/CF_{3}Br:54/36/10$ mixture. 
A composition like this was in use for the beginning of RPC development and allow
as to obtain a rich picture of the streamer process.

\section{Experimental results}
A first part of this section presents a behavior of induced charges in
the avalanche to streamer   transition region. The  second  subsection
shows  the data on a  spectrum of the streamer  light  coming from the
RPC. Examples of optical images are in the third part.

\subsection{Avalanche to streamer transition}
\begin{figure}                                                
\epsfxsize15cm\epsfysize13cm\epsffile{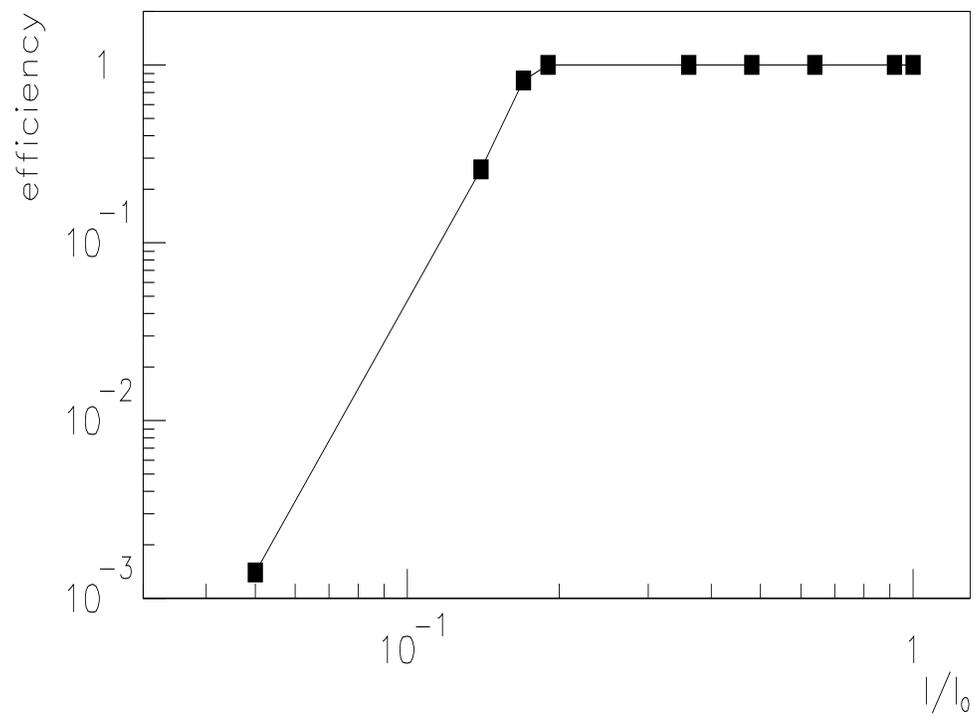}
\caption{\label{2} Probability to get the RPC reply at
different intensities of the laser pulse. $HV=7.8~kV$.}
\end{figure}

First of  all, we had  looked  how efficiency ($\epsilon$) of  the RPC
excitement  depends  on the  laser  pulse intensity ($I$). Fig.\ref{2}
shows $\epsilon$ versus $I/I_0$, where  $I_0$ is the maximal intensity
what we could get with  the LGI-21. This result  was obtained with the
voltage, $HV=7.8~kV$,  at which no streamer  signal was  observed when
our RPC  was tested on a   cosmic. As fig.\ref{2} demonstrates  that a
reply   from the  RPC   is always seen   at  $I/I_0~>~0.2$. Below this
intensity not every UV pulse produce the discharge.  Only one of $\sim
10^3$  pulses evokes discharge at $I/I_{0}\le  0.05$. We think that at
this low intensity, $\sim 5\%$ of  $I_0$, UV light occasionally knocks
out no more than one electron from cathode.   This assumption based on
a observation of almost $100\%$ efficiency for  the our chamber on the
cosmic when  avalanche  could  start from  a one   primary  ionization
cluster.  Tests with low  intensity are called  here as a work in "one
electron" mode.

\par
\begin{figure}
\epsfxsize15cm\epsfysize18cm\epsffile{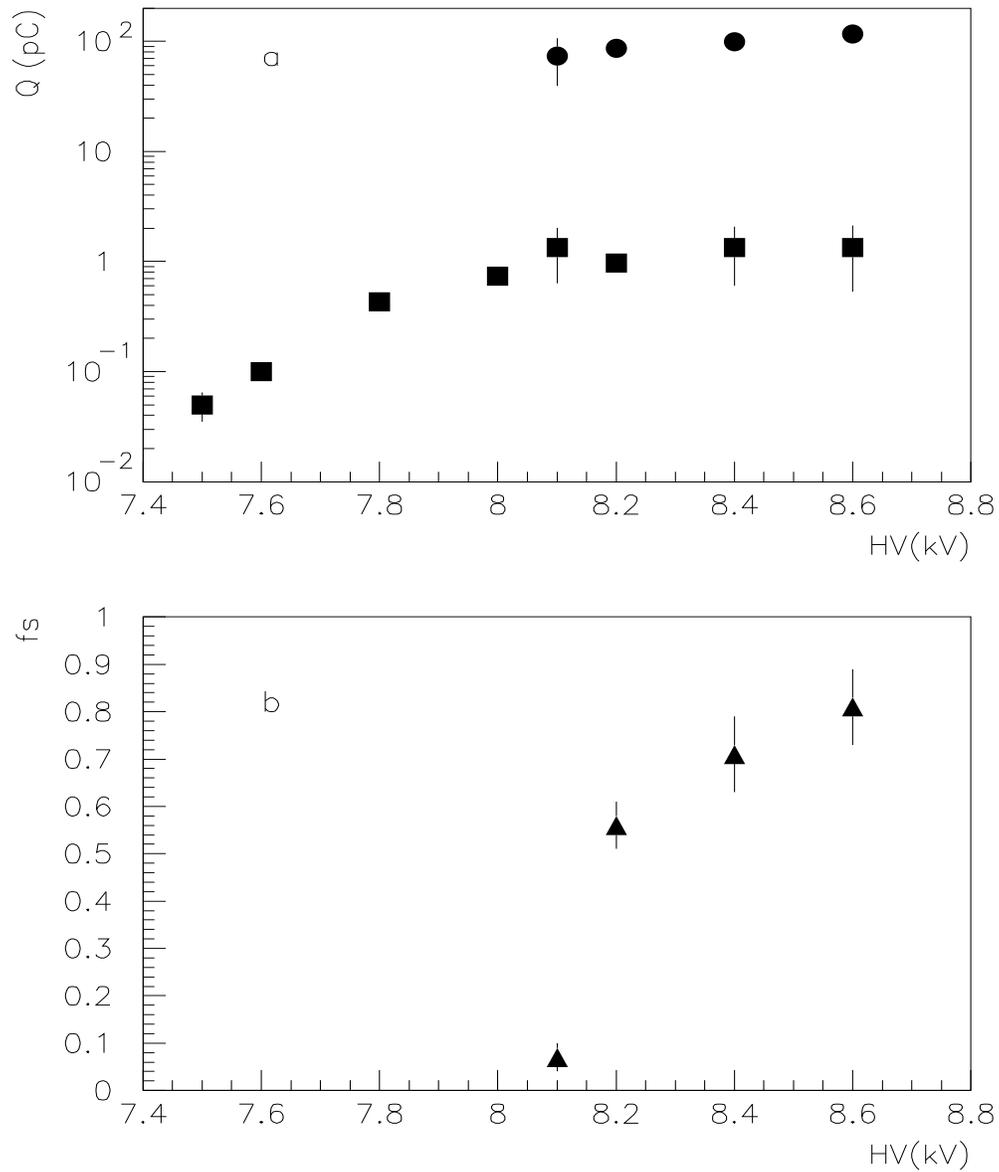}
\caption{\label{3} Measured at different high voltages: a~-~mean charge of the avalanche (boxes) and
streamer (circles), b~-~fraction of the 'streamer-A' signal.
The data was obtained in the "one electron" mode.}  
\end{figure}

 
A mean  value of the induced charge  ($<Q>$) was measured  in the
"one electron" mode and is shown in fig.\ref{3}a  as a function of the
high voltage. A  behavior of $<Q>$ in fig.\ref{3}a  looks like what we
saw with this gas mixture in cosmic tests: the avalanche signal rising
from  $0.05~pC$ to $\sim 1~pC$  with the $HV$  growth from $7.5~kV$ to
$\sim 8.2~kV$  and becomes to  be accompanied  by  the strong streamer
signal ($\sim  100~pC$)   at $HV~>~8.1-8.2~kV$.   A   fraction of  the
streamer at  different $HV$  is  given in fig.\ref{3}b.  The avalanche
signal caused by the laser beam has a less  variation in the amplitude
than the charged particle does. It  is so because the avalanche growth
always  starts  by the  electron knocked out   from the cathode and no
variation in the primary charge  position along the  gas gap as it  is
for the case  of crossing particle. That  is why the avalanche  signal
can be easily separated from the noise  (pedestal) even at $<Q>\approx
0.05~pC$.  A  behavior of  the   streamer induced signal in   the "one
electron" mode looks like a  set of pulses  (sparks) following in time
one  by one.   The  delay between the avalanche   signal and the first
streamer pulse can be up to  $100~ns$. Below we call  this type of the
streamer as a "streamer-A".

\par
\begin{figure}
\epsfxsize15cm\epsfysize18cm\epsffile{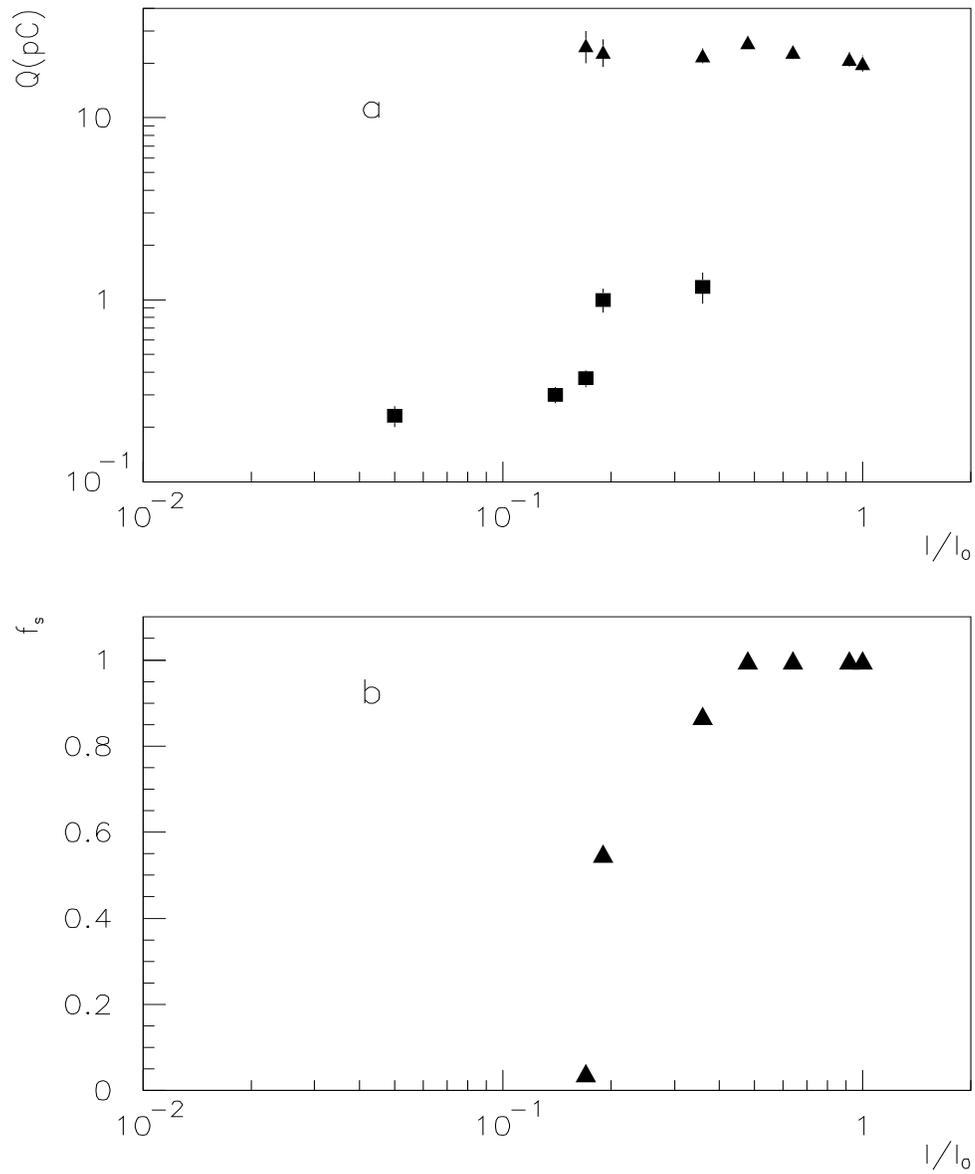}
\caption{\label{4} A top picture
shows charges of the avalanche (boxes) and "streamer-B" (triangles)
as a function of the light intensity. The fraction of the "streamer-B"
at different $I/I_0$ is given in the bottom picture. The data was
obtained at HV=7.8~kV.}
\end{figure}

 
It  was interesting for us to  see how the induced signal changes
if to increase a number of knocked out primary electrons. Fig.\ref{4}a
shows the $<Q>$ as a function of the ratio  $I/I_0$. A measurement was
carried out at the $HV=7.8~kV$.  The avalanche charge ($<Q_a>$, boxes)
rises  with a growth of  the  UV intensity. When  the $<Q_a>$  reaches
about  $1~pC$ the streamer   signal (triangles) appear. However,  this
streamer  signal differs from the   "streamer-A" observed at more high
$HV$  in the "one electron"  mode. This new  streamer consists of only
one pulse  with the charge   of  $30~pC$.  Its amplitude  spectrum  is
rather narrow. The distribution width  is few $pC$ only. This streamer
follows the initial avalanche with the few $ns$ delay and its duration
is $7-10~ns$(FWHM). We call this  streamer as a "streamer-B".  In  the
range of $I/I_{0}=0.2-1$ no considerable variation in the "streamer-B"
charge was found.  Fraction of the "streamer-B"  at the different $HV$
is presented in fig.\ref{4}b.

\par
\begin{figure}
\epsfxsize15cm\epsfysize9cm\epsffile{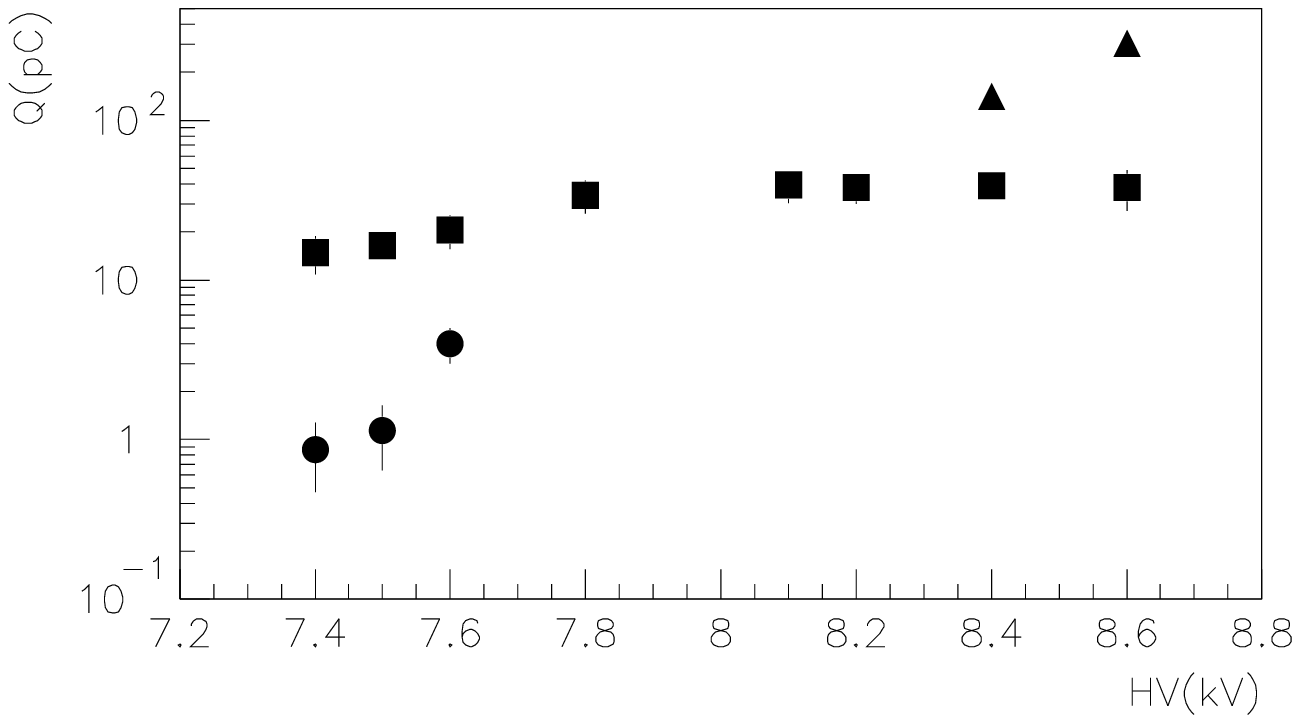}
\caption{\label{5} Three types of the signal observed at different $HV$.
The intensity of the light $I/I_0\approx 1$. The avalanche charge is given with circles, the "streamer-A" is shown with triangles,
black boxes describe the "streamer-B".}
%\end{figure}

%%%%%%%%%%%%%%%%%%% Fig.6    probabilily: str1 and str2
%\begin{figure}
\epsfxsize15cm\epsfysize9cm\epsffile{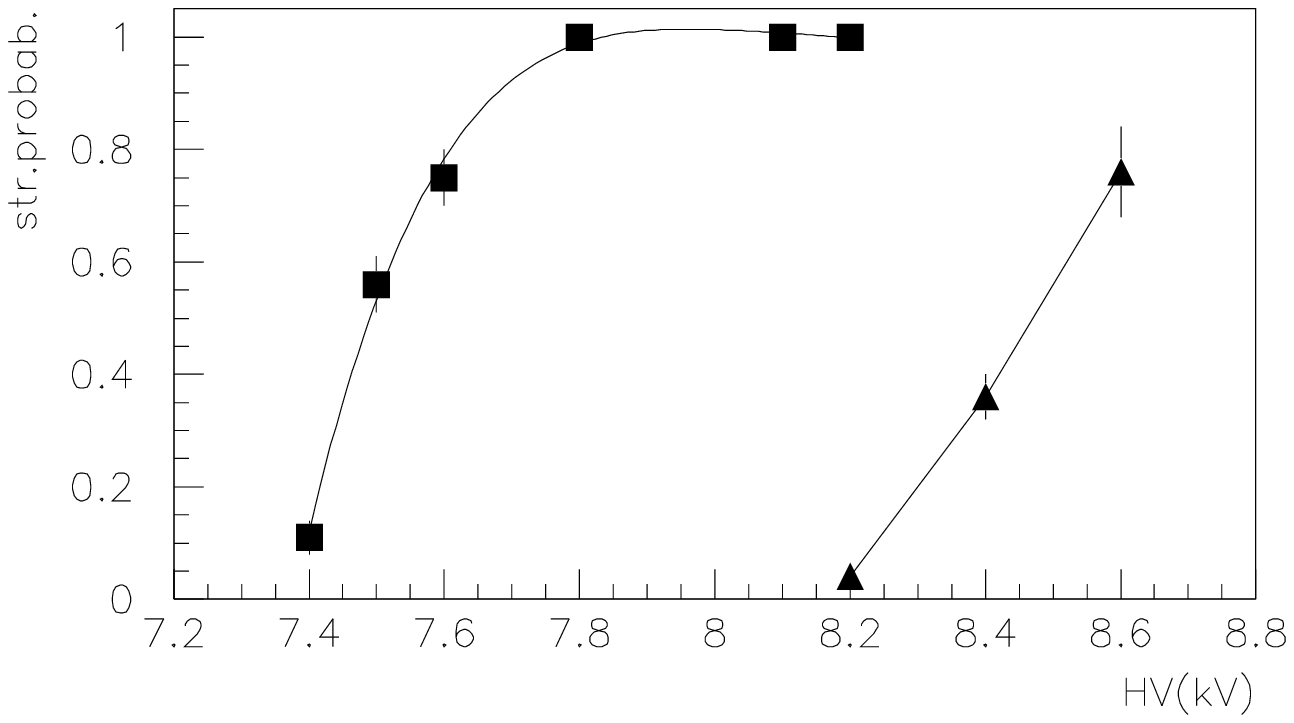}
\caption{\label{6} A probability of the "streamer-A" (triangles) and
"streamer-B" (boxes)  at different $HV$.
The UV light intensity is maximal.}
\end{figure}

 
At the  next  step we looked  for  signals  from the RPC   at the
maximal intensity  of  the UV beam.  Fig.\ref{5}  gives a mean induced
charge as a  function of the $HV$  when the $I/I_{0}$ was $\approx 1$.
Below $7.4~kV$ only avalanche signal was  observed. The mean avalanche
charge is  shown  in the figure  with circles.  As fig.\ref{5} show at
$7.4~kV$ the  streamer     signal   (boxes) with    a    $<Q_s>=15~pC$
appears. This streamer signal is the "streamer-B": it consists of only
one narrow (FWHM$\approx 10~ns$) pulse. The variation in its amplitude
is  about  $10-15\%$. The   time  between  the   "streamer-B" and  the
avalanche signal is few $ns$.  The amplitude of the "streamer-B" rises
slowly with $HV$ from $15~pC$ to about  $\sim 30~pC$. A probability to
observe "streamer-B" at    different  $HV$ is shown   in   fig.\ref{6}
(boxes). At voltages of $7.8-8.3~kV$  "streamer-B" is only signal what
we  could see.  Starting from  $HV\sim 8.3~kV$  both  "streamer-B" and
"streamer-A" (triangles) can be observed. Fraction of the "streamer-A"
rises with $HV$ as it is shown in  fig.\ref{6} (triangles). A presence
of the "streamer-A" can be noticed by  eye through the RPC glass wall:
 it looks like a set of sparks  near the point where light enters
the RPC. Some of sparks are few $cm$ far from this point.

\subsection{Light coming from discharges}
\begin{figure}
\epsfxsize15cm\epsfysize9cm\epsffile{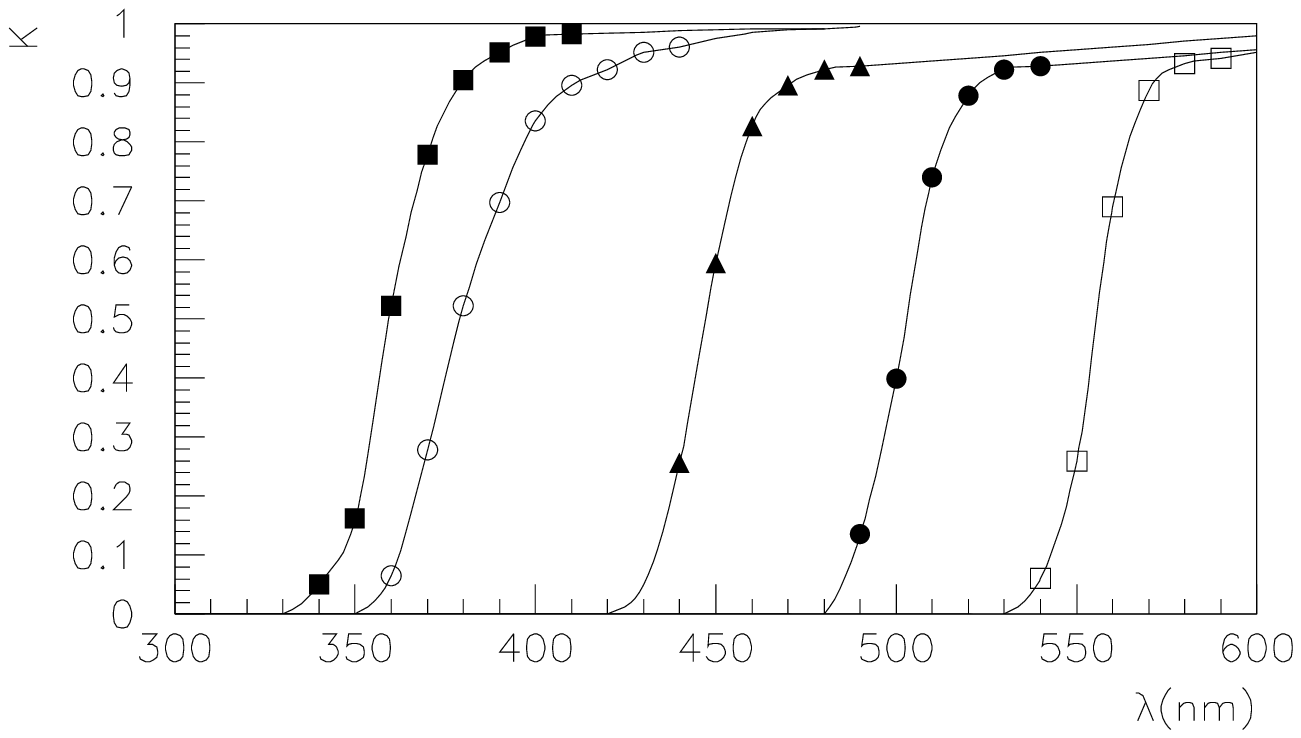}
\caption{\label{7} A transparancy as a function of the wavelength for five optical
filters: BS8(black boxes), JS4(sweet circles), JS12(triangles),
JS18(black circles), OS13(sweet boxes). Curves are taken from
specifications given by manufacturer of filters.}
%\end{figure}

%%%%%%%%%%%%%%%%%%% Fig.8 
%\begin{figure}
\epsfxsize15cm\epsfysize9cm\epsffile{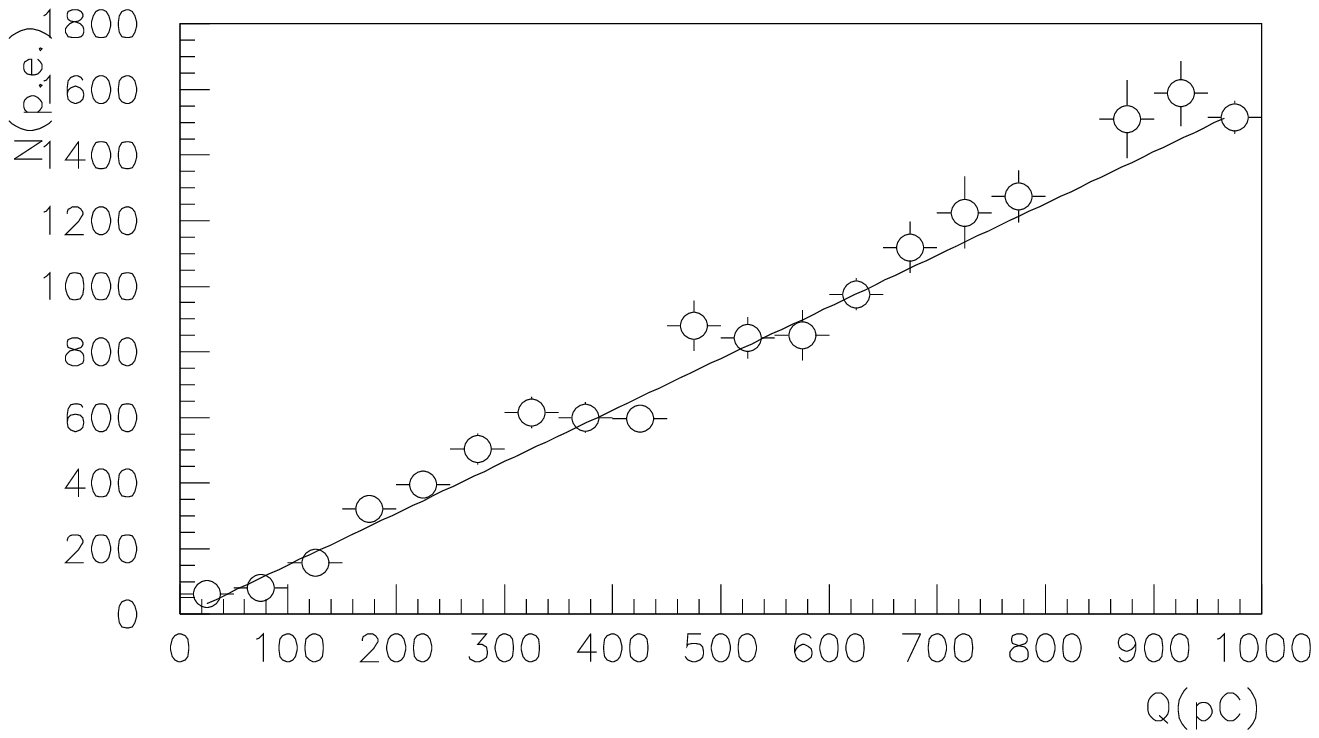}
\caption{\label{8} The number of photoelectrons (light from the streamer) as a function
of the induced charge. There is no filter between PM and RPC. A line shows
the result of the fit by linear law.}
\end{figure}

A  light coming   from  discharges through  the   glass electrode  was
measured with a  photomultiplier "FEU-87".  To  estimate a spectrum of
the discharge light  a set of measurements was  carried out  with five
different optical filters (BS8,  JS4, JS12, JS18 and  OS13), installed
between the RPC  and photocathode. A transparency  of these filters at
different   wave  length is given   in   fig.\ref{7}.

\par 
From  the
beginning we   should say  that  attempts  to measure  light  from the
avalanche  were failed. The avalanche light  is  rather week. It is at
level of  a  background  light. The  background   light appear  due to
reemission by the glass (and by the  dust on the glass surface) during
few ten   $ns$  after  UV   pulse.  As  for   the   streamer a    good
proportionality  between the RPC    induced charge and the number   of
photoelectrons ($N_{p.e.}$) was observed. The value of $N_{p.e.}$ as a
function of the $<Q>$ is given in fig.\ref{8}. The data in this figure
were obtained without any optical filter between the RPC  and PM.  The
figure demonstrates a  good proportionality between induced charge and
light output. A linear approximation of the data  brings a scale value
of  $\sim 1.59\pm 0.02~p.e./pC$.  By the  same  way, $N_{p.e.}$ versus
$<Q>$  was gotten  for cases   when PM was    added with five  filters
mentioned above. Scales  found after approximation  with a  linear law
are presented in the following table for the each filter. 

\small
\begin{tabular}{|r|c|c|c|c|c|}
\hline

filter & BS8 & JS4 & JS12 & JS18 & OS13 \\   \hline

$N_{p.e.}/pC$ &$1.56\pm 0.04$ &$1.46\pm 0.02$ &$0.94\pm 0.02$ 
&$0.55\pm 0.01$ & $ 0.12\pm 0.002$ \\
\hline 
\end{tabular}

\begin{figure}
\epsfxsize15cm\epsfysize9cm\epsffile{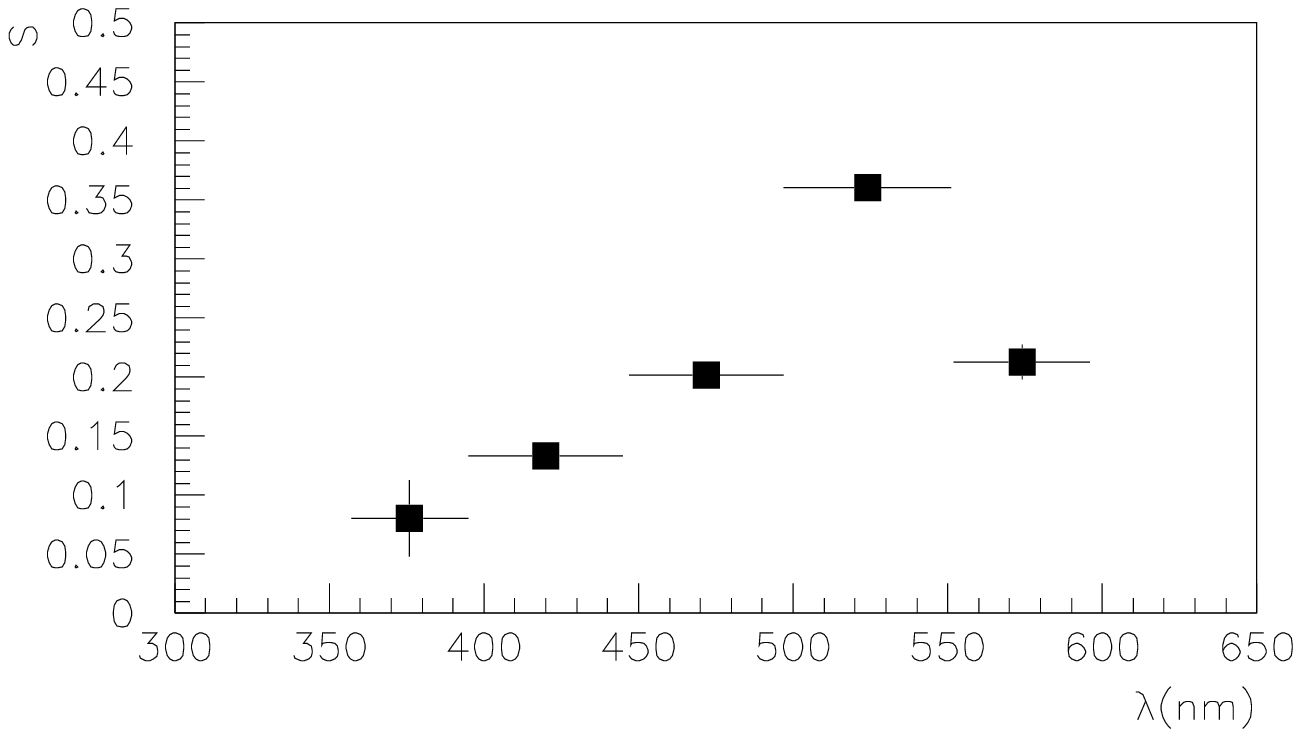}
\caption{\label{9} A reconstructed spectrum of the streamer light. It is normalized by unit.}
%\end{figure}
%%%%%%%%%%%%%%%%%% Fig.10
%\begin{figure}
\epsfxsize15cm\epsfysize9cm\epsffile{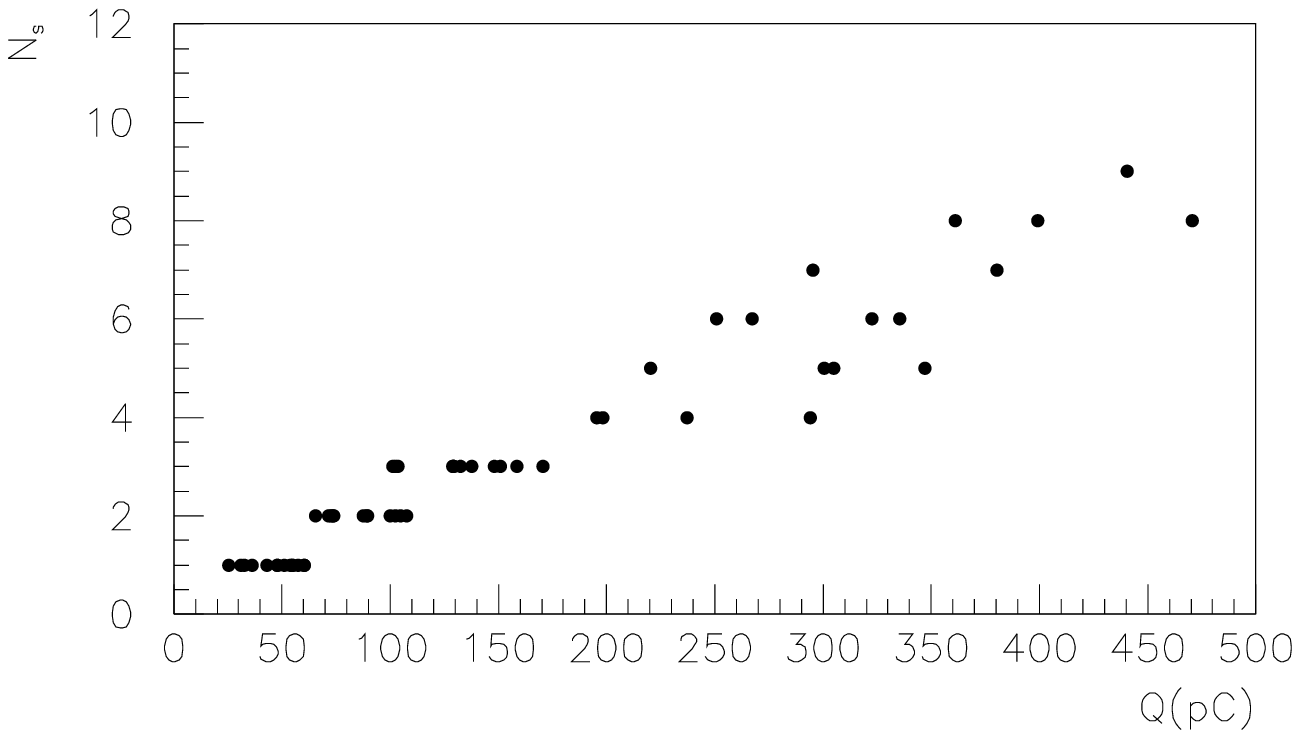}
\caption{\label{10} A number of light spots as a function of the induced charge.}
\end{figure}

\vskip 0.5cm 
\normalsize 
The table  gives a possibility to reconstruct
the spectrum of the streamer light coming through the glass cathode in
$\lambda\approx   300-600~nm$   range.   Corrected    for  the  FEU-87
sensitivity characteristic this spectrum  is shown in  fig.\ref{9}. As
can be seen the spectrum fast  decrease as the $\lambda$ decrease from
$\sim 500~nm$. It   could be result  of  the light  absorption  in the
window glass for  the  $\lambda \le  500~nm$.  A  range of  registered
wavelengths correspond to the photon energy of $2-4~eV$.
  
\subsection{Images of the streamer discharge}
As was said above the streamer looks by eye through the glass electrode as one or several bright spots. 

To get images  of discharges we used  the same glass RPC  as it was in
measurements  of  the light spectrum. The  RPC  was  viewed with Image
Intensifier ($II$)    "KANAL" having   bialkali photocathode.   A high
voltage pulse  sent  to a microchannel plate  inside  "KANAL" open the
Image Intensifier.
 
A  output screen  of  the  $II$ was  viewed   with the CCD  camera.  A
self-made card inside the IBM  PC was used  to digitize output signals
from the  CCD  camera. Images were  recorded into  files. Furthermore,
images were   observed during the data taking   on a computer monitor.

\par 
It was  found that the streamer  looks  through the glass  window
like one or  several bright spots. The  diameter of each spot is about
$2~mm$.  Fig.\ref{10} supports a idea  that a  correlation between the
'optical' and 'electrical'  information  exists.  This figure  show  a
number of   bright  spots ($N_s$) observed  through  the   glass as  a
function of the induced charge.  A good enough proportionality between
the $N_s$ and  charge can  be seen  in  the figure.  
\par
\begin{figure}
\epsfxsize14cm\epsfysize8cm\epsffile{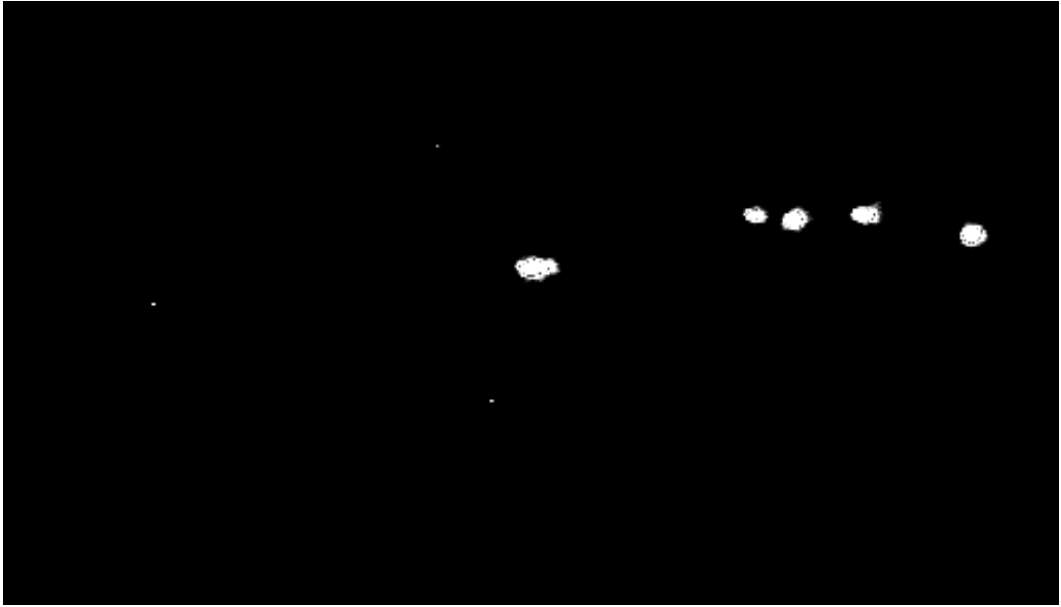}
\vskip 1cm
\epsfxsize14cm\epsfysize8cm\epsffile{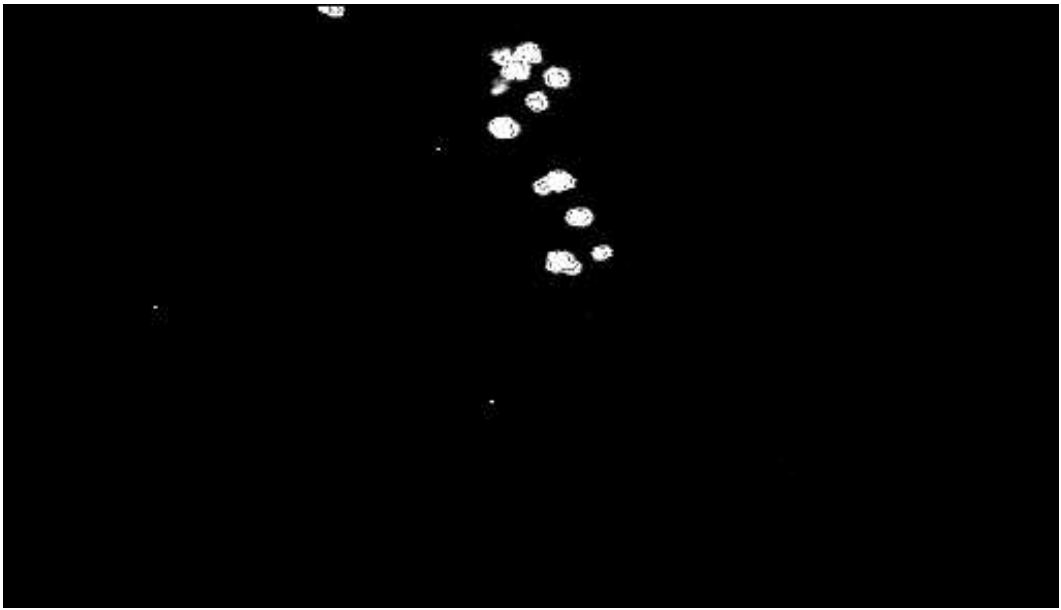}
\caption{\label{11} Two examples of picture obtained with CCD 
camera for the streamer discharge.}
\end{figure}

 
As  examples showing how the streamer develops     in the space, figure  \ref{11}
presents  two pictures obtained with the  $II$ and CCD  camera. A real
size of the each image is $3.5\times 5.5~cm^2$. A spot position, which
is the most close to  the center of  the  picture, corresponds to  the
place where UV  beam crossed the  RPC.  Pictures were obtained for the
"streamer-A" process.

\section{Conclusions}
\par  
The excitement of the  glass RPC by  the UV pulse laser provided
with interesting information on  how the avalanche transforms into the
streamer.  Two types of the streamer  have been observed. One of them,
"streamer-B", is produced  by the field of the  primary avalanche - it
appears    when the avalanche   charge (exactly   say, induced charge)
reaches  $\sim 1~pC$  and  it weakly depends   on the initial electric
field in   the   wide range  of $HV$.  The   "streamer-B"  follows the
avalanche discharge with few $ns$ delay only.

\par 
The second  type  of streamer,  "streamer-A", appears  when  $HV$
exceeds     some value.    For    our   gas   mixture   it    was   at
$HV~>~8.2~kV$.  Appearance of the  "streamer-A" does not depend on the
primary electrons  number. By  eye, this  streamer looks like  several
sparks  round point  where the laser  light   enters the RPC.  Some of
sparks are few $cm$ far from the primary  point. As it was said in \cite{sem} 
it seems this kind of
process in the RPC is connected with  a secondary photon emission from
primary   discharge.   Secondary  discharges     should  start    from
photoelectrons knocked out from    cathode  by photons from  a   short
wavelength  tail  of  the emission  spectrum.  Because  there is  weak
electric field  around  the primary discharge  the secondary discharge
can start only at some distance from initial.

\par 
The  estimations of the  streamer light spectrum and intensity of
the light from  discharges in the RPC  have  been done.  We hope these
data can  be  useful in   the future  for  everybody who  will try  to
reconstruct a space-time  picture of processes in the  RPC by use  the
optical information.

\newpage

%%%%%%%%%%%%%%%%%% Fig.1    set-up

%%%%%%%%%%%%%%%%%%  Fig.2     laser eff

%%%%%%%%%%%%%%%%%% Fig.3     mean charge

%%%%%%%%%%%%%%%%% Fig.4         1e mode

%%%%%%%%%%%%%%%%% Fig.5          3 types of signal

%%%%%%%%%%%%%%%%%%% Fig.7  transparancy of filters
%%%%%%%%%%%%%%%%%% Fig.9

%%%%%%%%%%%%%% Images

\end{document}